\documentclass[a4paper,amsmath,amssymb,10pt,aps,pre,twocolumn,floats,floatfix,superscriptaddress,showpacs]{revtex4}

\usepackage{amssymb,epsf}
\usepackage{graphicx} 
\usepackage{dcolumn} 
\usepackage{bm}
\usepackage{epsfig}
\usepackage{color}

\begin{document}

\title{Vulnerability of weighted networks}

\author{Luca Dall'Asta}
\affiliation{Laboratoire de Physique Th\'eorique (UMR du CNRS 8627),
B\^atiment 210, Universit\'e de Paris-Sud, 91405 ORSAY Cedex (France)}
\author{Alain Barrat}
\affiliation{Laboratoire de Physique Th\'eorique (UMR du CNRS 8627),
B\^atiment 210, Universit\'e de Paris-Sud, 91405 ORSAY Cedex (France)}
\author{Marc Barth{\'e}lemy}
\affiliation{School of Informatics and Department of
Physics, Indiana University, Bloomington, IN 47408, USA}
\affiliation{CEA-D\'epartement de Physique Th{\'e}orique
et Appliqu{\'e}e, 91680 Bruyeres-Le-Chatel, France}
\author{Alessandro Vespignani}
\affiliation{School of Informatics and Department of
Physics, Indiana University, Bloomington, IN 47408, USA}

\begin{abstract}
In real networks complex topological features are often associated
with a diversity of interactions as measured by the weights of the
links. Moreover, spatial constraints may as well play an important
role, resulting in a complex interplay between topology, weight, and
geography.  In order to study the vulnerability of such networks to
intentional attacks, these attributes must be therefore considered
along with the topological quantities. In order to tackle this issue,
we consider the case of the world-wide airport network, which is a
weighted heterogeneous network whose evolution and structure are
influenced by traffic and geographical constraints. We first
characterize relevant topological and weighted centrality measures and
then use these quantities as selection criteria for the removal of
vertices. We consider different attack strategies and different
measures of the damage achieved in the network. The analysis of
weighted properties shows that centrality driven
attacks are capable to shatter the network's communication or
transport properties even at very low level of damage in the 
connectivity pattern. The inclusion of weight and traffic therefore 
provides evidence for the extreme vulnerability of complex 
networks to any targeted strategy and need to be considered as key
features in the finding and development of defensive strategies.
\end{abstract}

\maketitle

\section{Introduction}
The network representation applies to large communication
infrastructure (Internet, e-mail networks, the World-Wide-Web),
transportation networks (railroads, airline routes), biological
systems (gene and/or protein interaction networks) and to a variety of
social interaction
structures~\cite{bara02,mendes03,psvbook,AmaralPNAS}. Very
interestingly, many real networks share a certain number of
topological properties. For example, most networks are
small-worlds~\cite{watts98}: the average topological distance between
nodes increases very slowly (logarithmically or even slower) with the
number of nodes.  Additionally, ``hubs'' [nodes with very large degree $k$
compared to the mean of the degree distribution $P(k)$] are often
encountered. More precisely, the degree distributions exhibit in many
cases heavy-tails often well approximated for a significant range of
values of degree $k$ by a power-law behavior ($P(k) \sim
k^{-\gamma}$)~\cite{bara02,mendes03} from which the name scale-free
networks originated. Real networks are however not only specified by
their topology, but also by the dynamical properties of processes
taking place on them, such as the flow of information or the traffic
among the constituent units of the system. In order to account for
these features, the edges are endowed with weights: for example, the
air-transportation system can be represented by a weighted network, in
which the vertices are commercial airports and the edges are non-stop
passenger flights. In this context, a natural definition of link
weights arises, as the capacity (in terms of number of passengers) of
the corresponding flight. Data about real weighted networks
(communication and infrastructure networks, scientific collaboration
networks, metabolic networks, etc.)  have been recently studied,
giving particular attention to the relation between weight properties
and topological
quantities~\cite{barrat:2004a,almaas:2004,Li:2003a}. These findings
have also generated several studies concerning modeling approaches in
which the mutual influence of weights and topology plays an explicit
role in determining network's properties
\cite{barrat:2004b,barrat:2004c,bianconi,dorogovtsev:2004,wang:2005}.

One of the most striking effects of the complex topological features
of networks concerns their vulnerability to attacks and random
failures.  Compared to ``regular'' $d$-dimensional lattices and random
graphs with a bounded degree distribution, heavy-tailed networks can
tolerate very high levels of random
failure~\cite{havlin01,newman00}. On the other hand, malicious attacks
on the hubs can swiftly break the entire network into small
components, providing a clear identification of the elements which
need the highest level of protection against such attacks
\cite{barabasi00,pv01a}. In this context it is therefore important to
study how the introduction of traffic and geographical properties may
alter or confirm the above findings. In particular we are interested
in two main questions: (i) which measures are best suited to assess
the damage suffered by weighted networks and to characterize the most
effective attack (protection) strategies; (ii) how traffic and spatial
constraints influence the system's robustness. In this article, our
attention is therefore focused on weighted networks with geographical
embedding and we analyze the structural vulnerability with respect to
various centrality-driven attack strategies. In particular, we propose
a series of topological and weight-depending centrality measures that
can be used to identify the most important vertices of a weighted
network. The traffic integrity of the whole network depends on the
protection of these central nodes and we apply these considerations to
a typical case study, namely the world-wide airport network.  We find
that weighted networks are even more vulnerable than expected in that
the traffic integrity is destroyed when the topological integrity of
the network is still extremely high. In addition all attacks
strategies, both local and non-local perform with almost the same
efficacy. The present findings may help in providing a quantitative
assessment of the most vulnerable elements of the network and the
development of adaptive reactions aimed at contrasting targeted
attacks.\\

\section{Network data set}
 
In the following we use the world-wide air-transportation network
(WAN), built from the International Air Transportation Association
database (www.iata.org).  This database contains the direct flight
schedules and available seats data from the vast majority of the
world's airlines for the year 2002. 
The network obtained from the IATA database contains $N=3,880$ interconnected
airports (vertices) and $18,810$ direct flight connections
(edges). This corresponds to an average degree of $\langle k \rangle =
9.7$, while the maximal one is $k_{max}=318$ showing a strong
heterogeneity of the degrees. This is confirmed by the fact that the
degree distribution can be described by the functional form $P(k)\sim
k^{-\gamma}f(k/k_c)$, where $\gamma \simeq 2.0$ and $f(k/k_c)$ is an
exponential cut-off which finds its origin in physical constraints on
the maximum number of connections that can be handled by a single
airport~\cite{luisair,guimera}. The WAN is a small-world: the average
shortest path length, measured as the average number of edges
separating any two nodes in the network, is $\langle \ell \rangle =
4.4$. The data contained in the IATA database allow to go beyond the
simple topological representation of the airports connections by
obtaining a weighted graph~\cite{weightbook} 
that includes the traffic $w_{ij}$ and actual length
$d_{ij}$ of each link, specyfying respectively the number of available seats
in flights between cities $i$ and $j$ during the year 2002
and  the euclidean distance $d_{ij}$ 
specifying the route length between cities $i$ and
$j$~\cite{barrat:2004a,guimera,Barrat:2005}.
The weights are symmetric ($w_{ij}=w_{ji}$)
for the vast majority of edges so that we work with a symmetric
undirected graph.  In addition to the very large degree fluctuations, both
the weights and the strength are broadly 
distributed~\cite{barrat:2004a,Li:2003a} adding
another level of complexity in this network.

\section{Measures of centrality}

A key issue in the characterization of
networks is the identification of the most central nodes in the
system. Centrality is however a concept that can be quantified by
various measures. The degree is a first intuitive and local quantity
that gives an idea of the importance of a node. Its natural
generalization to a weighted graph is given by the strength of
vertices defined for a node $i$ as~\cite{yookwt,barrat:2004a}
\begin{equation}
  s_i=\sum_{j \in {\cal V}(i)} w_{ij}~,
\label{Eq_strength}
\end{equation}
where the sum runs over the set ${\cal V}(i)$ of neighbors of $i$.  In
the case of the air transportation network it quantifies the traffic
of passengers handled by any given airport, with both a broad
distribution and strong correlations with the degree, of the form
$s(k) \sim k^{\beta_s}$ with $\beta_s \approx 1.5$~\cite{barrat:2004a}
(a random attribution of weights would lead to $s\sim k$ and thus
$\beta_s=1$).

Since space is also an important parameter in this network, other interesting
quantities are the {\em distance strength} $D_i$ and {\em outreach}
$O_i$ of $i$
\begin{equation}
D_i = \sum_{j \in {\cal V}(i)} d_{ij}~,
\ \ O_i = \sum_{j \in {\cal V}(i)} w_{ij} d_{ij}~,
\end{equation}
where $d_{ij}$ is the {\em Euclidean} distance between $i$ and
$j$. These quantities describe the cumulated distances of all the
connections from the considered airport and the total distance
traveled by passengers from this airport, respectively.  They display
both broad distributions and grow with the degree as $D(k) \sim
k^{\beta_D}$ with $\beta_D \approx 1.5$~\cite{Barrat:2005}, and $O(k)
\sim k^{\beta_O}$, with $\beta_O \approx 1.8$, showing the existence
of important correlations between distances, topology and traffic.

Such local measures however do not take into account non-local
effects, such as the existence of crucial nodes which may have small
degree or strength but act as bridges between different part of the
network.  In this context, a widely used quantity to investigate node
centrality is the so-called betweenness centrality (BC)
\cite{freeman}, which counts the fraction of shortest paths between
pairs of nodes that passes through a given node. More precisely, if
$\sigma_{hj}$ is the total number of shortest paths from $h$ to $j$
and $\sigma_{hj}(i)$ is the number of these shortest paths that pass
through the vertex $i$, the betweenness of the vertex $i$ is defined
as $b_i=\sum_{h,j}\sigma_{hj}(i)/\sigma_{hj}$, where the sum is over
all the pairs with $j \neq h$.  Key nodes are thus part of more
shortest paths within the network than less important nodes.

In weighted networks, unequal link capacities make some specific paths
more favorable than others in connecting two nodes of the network. It
thus seems natural to generalize the notion of betweenness centrality
through a {\em weighted betweenness centrality} in which shortest
paths are replaced with their weighted versions.  
A straightforward way to generalize the hop distance (number of
traversed edges) in a weighted graph consists in assigning to each
edge $(i,j)$ a length $\ell_{ij}$ that is a function of the
characteristics of the link $i-j$. For example for the WAN, $\ell_{ij}$
should involve
quantities such as the weight $w_{ij}$ or the Euclidean distance
$d_{ij}$ between airports $i$ and $j$. It is quite natural to assume
that the effective distance between two linked nodes is a decreasing
function of the weight of the link:
the larger
the flow (traffic) on a path, the more frequent and the fastest will
be the exchange of physical quantities (e.g. information, people,
goods, energy, etc.). In other words, we consider that the
``separation'' between nodes $i$ and $j$ decreases as $w_{ij}$
increases. While a first possibility would be to define the 
length of an edge as
the inverse of the weight, $\ell_{i,j}=1/w_{ij}$, 
we propose to also take into account the geographical embedding of
the network, through the following definition:
\begin{equation}
\ell_{ij}=\frac{d_{ij}}{w_{ij}} \ .
\end{equation}
It is indeed reasonable to consider two nodes of the networks as
further apart if their geographical distance is larger, however a
large traffic allows to decrease the ``effective'' distance by
providing more frequent travel possibilities.

For any two nodes $h$ and $j$, the weighted shortest path between $h$
and $j$ is the one for which the total sum of the lengths of the edges
forming the path from $h$ to $j$ is minimum, independently from the
number of traversed edges.  We denote by $\sigma^{w}_{hj}$ the total
number of weighted shortest paths from $h$ to $j$ and
$\sigma^{w}_{hj}(i)$ the number of them that pass through the vertex
$i$; the weighted betweenness centrality (WBC) of the vertex $i$ is then
defined as
\begin{equation}
b^{w}_{i}=\sum_{h,j} \frac{\sigma^{w}_{hj}(i)}{\sigma^{w}_{hj}}~,
\label{wbc}
\end{equation}
where the sum is over all the pairs with $j \neq h$~\footnote{ As
already noted by Brandes, the algorithm proposed in
Ref.~\cite{brandes} can be easily extended to weighted graphs, using
in addition Dijkstra's algorithm \cite{dijkstra} which provides a way
to compute weighted shortest paths in at most ${\cal O}(E N)$ where
$E$ is the number of edges. }.  The weighted betweenness represents a
trade-off between the finding of ``bridges'' that connect different
parts of a network, and taking into account the fact that some links
carry more traffic than others. We note that the definition
(\ref{wbc}) is very general and can be used with any definition of the
effective length of an edge $\ell_{ij}$.\\

\noindent
\subsection*{Centrality measures correlations} 

The probability distributions of
the various definitions of centrality are all characterized by heavy
tailed distributions.  In addition a significant level of correlation
is observed: vertices that have a large degree have also typically
large strength and betweenness. When a detailed analysis of the
different rankings is done,however we observe that they do not coincide
exactly. For example, in the case of the WAN the most connected
airports do not necessarily have the largest betweenness
centrality~\cite{luisair,guimera,Barrat:2005}. Large fluctuations between
centrality measures also appear when inspecting the list of the
airports ranked by using different definitions of centrality including
weighted ones: strikingly, each definition provides a different
ranking. In addition, some airports which
are very central according to a given definition, become peripheral
according to another criteria. For example, Anchorage has a large
betweenness centrality but ranks only $138^{th}$ and $147^{th}$ in terms
of degree and strength, respectively. Similarly, Phoenix or Detroit
have large strength but low ranks ($>40$) in terms of degree and
betweenness.

While previous analysis have focused on the quantitative correlations
between the various centrality measures here we focus on ranking differences
according to the various centrality measures.
A quantitative analysis of the correlations between two rankings of
$n$ objects can be done using rank correlations such as Kendall's
$\tau$~\cite{num_rec}
\begin{equation}
\tau=\frac{n_c-n_d}{n(n-1)/2}
\end{equation}
where $n_c$ is the number of pairs whose order does not change 
in the two different lists and $n_d$
is the number of pairs whose order was inverted. This quantity is
normalized between $-1$ and $1$:
$\tau=1$ corresponds to identical ranking while $\tau=0$
is the average for two uncorrelated rankings and $\tau=-1$ is a
perfect anticorrelation.
\begin{center}
\begin{table}[t]
\begin{tabular}{lccccccc}
\hline
  & $k$  & $D$ & $s$  &  $O$ &$BC$ & $WBC$ &\\
 \hline
Degree   $k$           &  1   & 0.7&   0.58  & 0.584  & 0.63 & 0.39 & \\ 
Distance strength $D$   & 0.7  & 1  &  0.56    & 0.68    & 0.48 & 0.23 & \\ 
Strength       $s$      & 0.58 & 0.56   &  1   & 0.83 &  0.404 & 0.24 & \\ 
Outreach   $O$    & 0.584 & 0.68 & 0.83 & 1  &  0.404  & 0.21 & \\ 
Betweenness     $BC$  & 0.63 &  0.48    &  0.404 & 0.404 & 1& 0.566& \\ 
Weighted $BC$  & 0.39 &  0.23 &  0.24 & 0.21 & 0.566 & 1 & \\ 
\hline
 \end{tabular}
\caption{Similarity between the various rankings as measured
by Kendall's $\tau$. For random rankings of $N$ values, the typical
$\tau$ is of order $10^{-2}$.}
\label{tab2}
\end{table}
\end{center}
Table \ref{tab2} gives the values of $\tau$ for all the possible
pairs of centrality rankings. For $N=3,880$, two random rankings
yield a typical value of $\pm 10^{-2}$ so that even the smallest
observed $\tau=0.21$ is the sign of a strong correlation (All the
values in this table were already attained for a sublist of only
the first $n$ most
central nodes, with $n\approx 500$). 
Remarkably enough, even a highly non-local 
quantity such as the BC is strongly correlated with the simplest local, non
weighted measure given by the degree. The weighted betweenness is the
least correlated with the other measures (except with the betweenness),
because $\ell_{ij}$ involves ratios of weights and distances.

\begin{figure} 
\centerline{
\includegraphics*[angle=-90,width=0.40\textwidth]{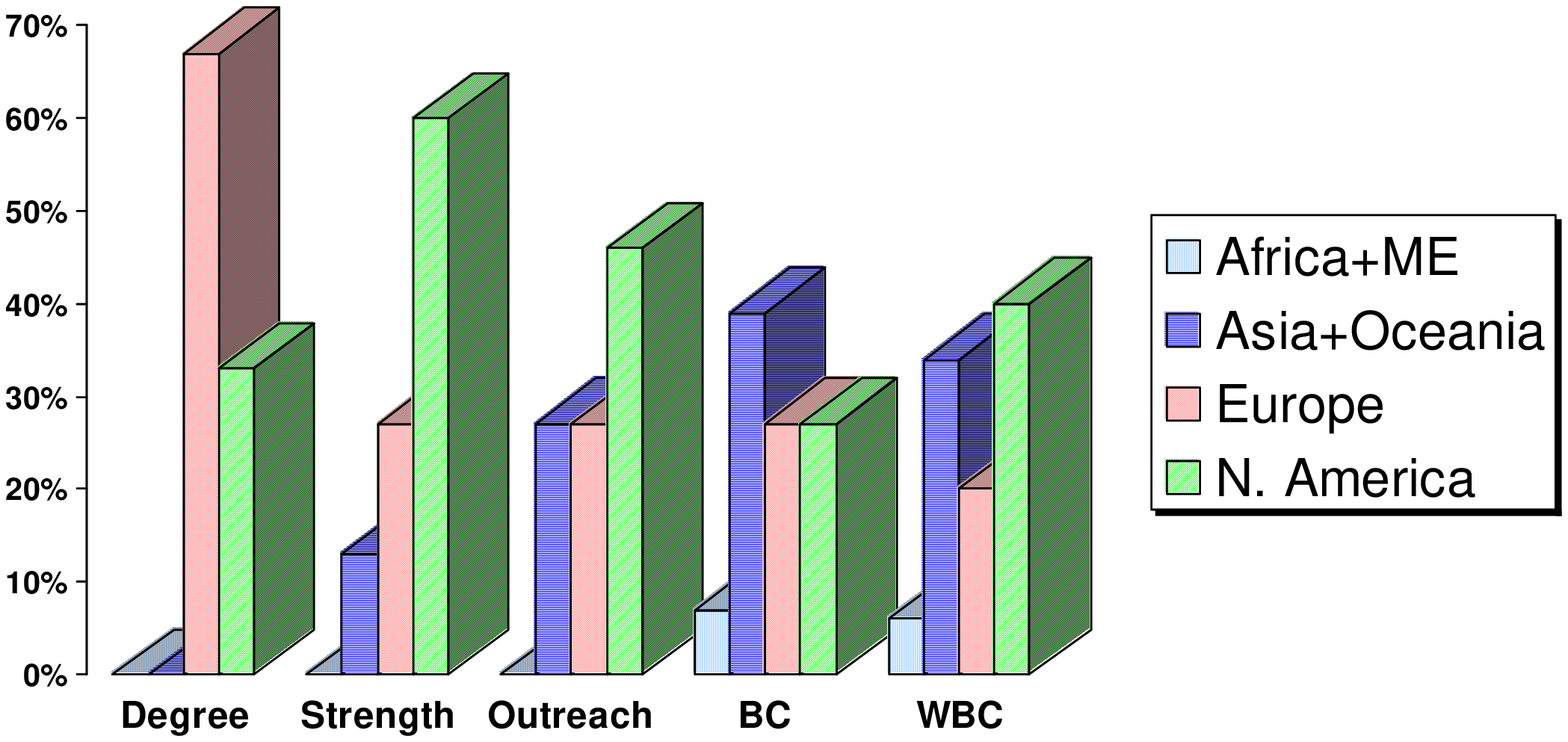}}
\caption{Geographical distribution of the world's $15$ most central
airports ranked according to different centrality measures. 
Topological measures miss the
economical dimension of the world-wide airport. In contrast, the
traffic aspect shows a clear dominance of North-America. Non-local
measures pinpoint important nodes in each geographical zone.}
\label{fig:geo}
\end{figure}

Another important issue concerns how the centrality ranking relates to
the geographical information available for infrastructure networks
such as the WAN.  Figure~\ref{fig:geo} displays the geographical
distribution of the world's fifteen most central airports ranked
according to different centrality measures. This figure highlights the
properties and biases of the various measures: on one hand,
topological measures miss the economical dimension of the world-wide
airport while weighted measures reflect traffic and economical
realities. Betweenness based measures on the other hand pinpoint the
most important nodes in each geographical zone. In particular, the
weighted betweenness appears as a balanced measure which combines
traffic importance with topological centrality, leading to a more
uniform geographical distribution of the most important nodes.\\

\section{Vulnerability of weighted networks}
\subsection{Damage Characterization} 
The example of the WAN enables us to
raise several questions concerning the vulnerability of weighted
networks. The analysis of complex networks robustness has indeed been
largely investigated in the case of unweighted
networks~\cite{barabasi00,havlin01,newman00,holme02}.  In particular,
the topological integrity of the network $N_g/N_0$ has been studied,
where $N_g$ is the size of the largest component after a fraction $g$
of vertices has been removed and $N_0$ is the size of the original
(connected) network. When $N_g \simeq \mathcal{O}(1)$, the entire
network has been destroyed \footnote{Since the topological integrity
focuses only on the largest component and overlooks the connectivity
of smaller components, one can also monitor the average inverse
geodesic length, also called efficiency~\cite{crucitti}}.

Damage is generally studied for increasingly larger fractions $g$ of
removed nodes in the network, where the latter are chosen following
different strategies. Heterogeneous networks with a scale-free degree
distribution are robust to situations in which the damage affects
nodes randomly. On the other hand, the targeted destruction of nodes
following their degree rank is extremely effective, leading to the
total fragmentation of the network at very low values of
$g$~\cite{havlin01,newman00,barabasi00}. Moreover, the removal of the
nodes with largest betweenness typically leads to an even faster
destruction of the network~\cite{holme02}.

In the case of weighted networks, the quantification of the damage
should consider also the presence of weights. In this perspective, the
largest traffic or strength still carried by a connected component of
the network is likely an important indicator of the network's
functionality. For this reason, we define new measures for the
network's damage
\begin{equation}
I_s(g)=\frac{{\cal S}_g}{{\cal S}_0}, \ \ \
I_O(g)=\frac{{\cal O}_g}{{\cal O}_0}~,\ \ \ 
I_D(g)=\frac{{\cal D}_g}{{\cal D}_0} ~,
\end{equation}
where ${\cal S}_{0}=\sum_i s_i$, ${\cal O}_0=\sum_i O_i$ and
${\cal D}_{0}=\sum_i D_i$ are the total strength, outreach
and distance strength in the undamaged network and
${\cal S}_g=\mbox{max}_G\sum_{i\in G} s_i$,
${\cal O}_g =\mbox{max}_G\sum_{i\in G} O_i$ 
and ${\cal D}_g=\mbox{max}_G\sum_{i\in G} D_i$
correspond to the largest strength, outreach or distance strength 
carried by any connected component $G$ in the network, after
the removal of a density $g$ of nodes. These quantities measure
the {\em integrity} of the network with respect to either
strength, outreach or distance strength, since they refer to the
relative traffic or flow that is still handled in the largest operating
component of the network.\\

\begin{figure} 
\centerline{
\includegraphics*[angle=0,width=0.4\textwidth]{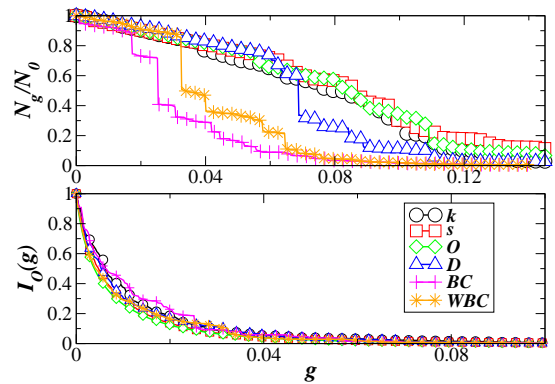}
}
\vspace*{.02cm}
\caption{Effect of different attack strategies on the size of the
connected giant component (top) and on the outreach (bottom).  }
\label{fig:damage}
\end{figure}

\subsection{Variable-ranking attack strategies}
In order to evaluate the vulnerability of the air-transportation
network WAN, we study the behavior of damage measures in the presence
of a progressive random damage and of different attack strategies. 
Similarly to the simple topological case, weighted networks are
inherently resilient to random damages. Even at a large
density $g$ of removed nodes, $N_g/N_0$ and all integrity measures
decrease mildly and do not seem to have a sharp threshold above which
the network is virtually destroyed. This is in agreement
with the theoretical prediction for the absence of a percolation
threshold in highly heterogeneous graphs
\cite{havlin01,newman00}. Very different is the scenario corresponding
to the removal of the most central nodes in the network. In this case,
however, we can follow various strategies based on the different
definitions for the centrality ranking of the most crucial nodes:
nodes can indeed be eliminated according to their rank in terms of degree,
strength, outreach, distance strength, topological betweenness, and
weighted betweenness. 
In addition, we consider attack strategies based on a recursive re-calculation of
the centrality measures on the network after each damage. This has
been shown to be the most effective strategy~\cite{holme02}, as 
each node removal leads to a change in the centrality properties 
of the other nodes. Such procedure is somehow akin to a
cascading failure mechanism in which each failure triggers a
redistribution on the network and changes the next most vulnerable
node.

In Fig.~\ref{fig:damage} we report the behavior of $N_g/N_0$ and of
the outreach integrity $I_O(g)$ for all cases. As expected, all
strategies lead to a rapid breakdown of the network with a very small
fraction of removed nodes.  More precisely, the robustness level of
the network depends on the quantity under scrutiny. First, the size of
the giant component decreases faster upon removal of nodes which are
identified as central according to global (i.e. betweenness)
properties, instead of local ones (i.e.  degree, strength), showing
that, in order to preserve the structural integrity of a network, it
is necessary to protect not only the hubs but also strategic points
such as bridges and bottle-neck structures. Indeed, the betweenness,
which is recomputed after each node removal is the most effective
quantity in order to pin-point such nodes~\cite{holme02}.  The
weighted betweenness combines shortest paths and weights and leads to
an intermediate result: some of the important topological bridges
carry a small amount of traffic and are therefore part of more
shortest paths than weighted shortest paths. These bridges have
therefore a lower rank according to the weighted betweenness. The
weighted betweenness is thus slightly less efficient for identifying
bridges. Finally, we note that all locally defined quantities yield a
slower decrease of $N_g$ and that the removal of nodes with the
largest distance strength is rather effective since it targets
nodes which connect very distant parts of the network.

Interestingly, when the attention shifts on the behavior of the
integrity measures, one finds a different picture in which all the
strategies achieve the same level of damage 
(the curves of $I_s(g)$ and $I_D(g)$ present shapes very 
close to the one of $I_O(g)$). Most importantly, their
decrease is even faster and more pronounced than for topological
quantities: for
$N_g/N_0$ still of the order of $80\%$, the integrity measures are typically
smaller than $20\%$. This emphasizes how the purely topological
measure of the size of the largest component does not convey all the
information needed. In other words, the functionality of the network 
can be temporarily jeopardized in terms of traffic even if the
physical structure is still globally well-connected.
This implies that weighted networks appear more fragile than thought
by considering only topological properties. All targeted strategies
are very effective in dramatically damaging the network, 
reaching the complete destruction at a very
small threshold value of the fraction of removed nodes. 
In this picture, the maximum damage is achieved still by strategies 
based on non-local quantities such as the betweenness which lead 
to a very fast decrease of both topological and traffic related
integrity measures. On the other hand, the results for the integrity
shows that the network may unfortunately
be substantially harmed also by using strategies based on
local quantities more accessible and easy to calculate. \\

\subsection{Single-ranking attack strategies.}
\begin{figure} 
\centerline{
\includegraphics*[angle=0,width=0.44\textwidth]{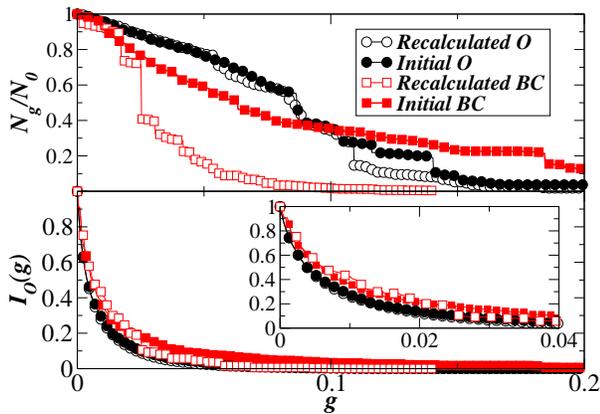}
}
\vspace*{.05cm}
\caption{Removal of nodes according to the ranking calculated at the
beginning of the process (empty symbols) or to recalculated rankings
(full symbols). The decrease of $N_g$ and $I_O(g)$ are
comparable for both cases. 
Inset: Initial decrease of $I_O(g)$ for very small values
of $g$.}
\label{fig:no_recalc}
\end{figure}

The previous strategies based on a recursive re-calculation of
the centrality measures on the network are
however computationally expensive and depend
upon a global knowledge of the effect of each node removal. It is
therefore interesting to quantify the effectiveness of such a strategy
with respect to the more simple use of the ranking information
obtained for the network in its integrity. In this case the nodes are
removed according to their initial ranking calculated for the
undamaged network. As shown in Fig.~\ref{fig:no_recalc}, successive
removals of nodes according to their initial outreach or BC lead to a
topological breakdown of the network which is maximized in the case
of recalculated quantities~\cite{holme02}. This effect is very clear
in the case of global measures of centrality  such as the betweenness
that may be altered noticeably by local re-arranegements. When traffic integrity
measures are studied, however, differences are negligible
(Fig.~\ref{fig:no_recalc}, bottom curves): a very fast
decrease of the integrity is observed for all strategies, based either
on initial or recalculated quantities.  The origin of the similarity
between both strategies can be traced back by studying how much the
centrality ranking of the network vertices is scrambled during the
damage process.  In order to quantify the reshuffling of the ranking
of the nodes according to various properties, we study the previously
used rank correlation as measured by Kendall's $\tau$, computed
between the rankings of the nodes according to a given property before
and after each removal. In all cases, $\tau$ remains very close to
$1$, showing that the reshuffling caused by any individual removal
remains extremely limited. Slightly smaller values are observed when
we compare the rankings of the betweenness or of the weighted
betweenness. This fact can be understood since such quantities are
non-local and the betweennesses is more prone to vary when any node in
the network is removed. This evidence brings both good and bad news
concerning the protection of large scale infrastructures. On one hand,
the planning of an effective targeted attack does need only to gather
information on the initial state of the network. On the other hand,
the identification of crucial nodes to protect is an easier task that
somehow is weakly dependent on the attack sequence.\\

\subsection{Geographical heterogeneoity.}
\begin{figure} 
\centerline{\includegraphics*[width=0.44\textwidth]{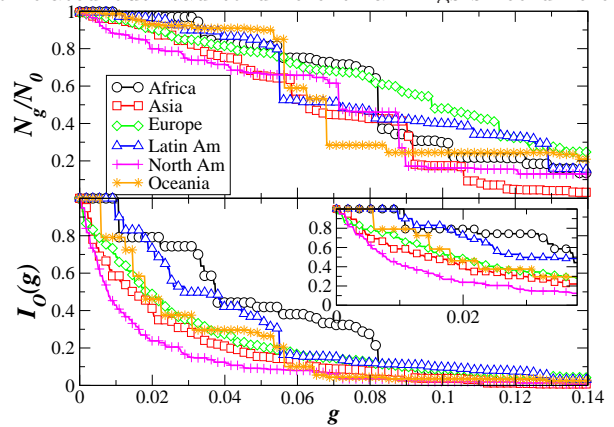}}
\vspace*{.05cm}
\caption{Geographical effect of the removal of nodes with largest
strength. The integrity decreases strongly in regions such as
North-America, while a ``delay'' is observed for the zones with
smaller initial outreach or strength.}
\label{fig:geo2}
\end{figure}

As shown in Fig.~\ref{fig:geo}, various geographical zones contain
different numbers of central airports. The immediate consequence is
that the different strategies for node removal have different impacts
in different geographical areas. Figure~\ref{fig:geo2} highlights this
point by showing the decrease of two integrity measures representative
of topological and traffic integrity, respectively. These quantities
were measured on subnetworks corresponding to the six following
regions: Africa, Asia, Europe, Latin and North America, and Oceania.
Figure~\ref{fig:geo2} displays the case of a removal of nodes
according to their strength (other removal strategies lead to similar
data). While the curves of topological damage are rather intertwined,
the decrease of the different integrity measures is much faster for
North America, Asia and Europe than Africa, Oceania and Latin America;
in particular the removal of the first nodes do not affect at all
these three last zones. Such plots demonstrate two crucial
points. First, various removal strategies damage differently the
various geographical zones. Second, the amount of damage according to
a given removal strategy strongly depends on the precise measure used
to quantify the damage. More generally, these results lead to the idea
that large weighted networks can be composed by different subgraphs
with very different traffic structure and thus different responses to
attacks.\\

\noindent
\section{Conclusions.}  In summary, we have identified a set of different
but complementary centrality measures for weighted networks. The
various definitions of centrality are correlated but lead to different
rankings since different aspects (weighted or topological, and local
or global) are taken into account. The study of the vulnerability of
weighted networks to various targeted attack strategies shows that
complex networks are more fragile than expected from the analysis of
topological quantities when the traffic characteristics are taken into
account. In particular, the network's integrity in terms of carried
traffic is vanishing significantly before the network is topologically
fragmented. Moreover, we have compared attacks based on initial
centrality ranking with those using quantities recalculated after each removal,
since any modification of the network (e.g. a node removal) leads to a 
partial reshuffling of these rankings. Strikingly, and in contrast to the case
of purely topological damage, the integrity of the network is harmed in a very
similar manner in both cases. All these results warn about the extreme
vulnerability of the traffic properties of weighted networks and signals the
need to pay a particular attention to weights and traffic in the 
design of protection strategies.\\

\noindent
{\bf Acknowledgments}

We thank IATA for making the airline commercial flight
database available.  A.B. and L.D. are partially supported by the EU
within the 6th Framework Programme under contract 001907
``Dynamically Evolving, Large Scale Information Systems'' 
(DELIS).



\begin{thebibliography}{10}

\bibitem{bara02}
A.-L. Barab\'{a}si, A.-L.  and  R. Albert,  
Rev. Mod. Phys. {\bf 74}, 47-97 (2002).

\bibitem{mendes03}
S. N. Dorogovtsev and J. F. F. Mendes,
{\it Evolution of Networks: from biological nets to the Internet and
WWW} (Oxford University Press, Oxford 2003).

\bibitem{psvbook}
R.~Pastor-Satorras and A.~Vespignani,
{\em Evolution and structure of the Internet: A statistical physics
approach} (Cambridge University Press, Cambridge, 2004).

\bibitem{AmaralPNAS} L.A.N.~Amaral, A.~Scala, M.~Barth{\'e}lemy, and
H.E.~Stanley, Proc. Nat. Acad. Sci. USA {\bf 97}, 11149 (2000).

\bibitem{watts98}
Watts D. J. and Strogatz S. H.,
Nature {\bf 393}, 440 (1998).

\bibitem{barrat:2004a} A.~Barrat, M.~Barth\'elemy, R.~Pastor-Satorras,
and A.~Vespignani, Proc. Natl. Acad. Sci. USA {\bf 101}, 3747 (2004).

\bibitem{almaas:2004} E. Almaas, B. Kov\'acs,
T. Viscek, Z. N. Oltvai and A.-L. Barab{\'a}si,
{\it Nature} {\bf 427}, 839 (2004).

\bibitem{Li:2003a} W.~Li and X. Cai, Phys. Rev. E {\bf 69}, 046106 (2004).

\bibitem{barrat:2004b} A.~Barrat, M.~Barth\'elemy, and A.~Vespignani,
Phys. Rev. Lett., {\bf 92}, 228701 (2004). 

\bibitem{barrat:2004c}
A.~Barrat, M.~Barth\'elemy, and A.~Vespignani,
Phys. Rev. E {\bf 70}, 066149 (2004).

\bibitem{bianconi}
G. Bianconi,
Europhys. Lett. {\bf 71}, 1029 (2005).

\bibitem{dorogovtsev:2004}
S. N. Dorogovtsev and J. F. F. Mendes, 
cond-mat/0408343.

\bibitem{wang:2005}
W.-X. Wang, B.-H. Wang, B. Hu, G. Yan and Q. Ou,
Phys. Rev. Lett. {\bf 94}, 188702 (2005).

\bibitem{havlin01} R. Cohen, K. Erez, D. ben-Avraham, and S. Havlin,
  Phys. Rev. Lett. {\bf 85}, 4626 (2000);

\bibitem{newman00} D. S. Callaway, M. E. J. Newman, S. H. Strogatz,
  and D. J. Watts, Phys. Rev. Lett. {\bf 85}, 5468 (2000).

\bibitem{barabasi00}
R. A. Albert, H. Jeong and A.-L. Barab{\'a}si, 
Nature {\bf 406}, 378 (2000).

\bibitem{pv01a} R. Pastor-Satorras, and A. Vespignani,
Phys. Rev. Lett. {\bf 86}, 3200 (2001).

\bibitem{luisair}
R.~Guimer\`a, S.~Mossa, A.~Turtschi, and
L.A.N.~Amaral, Proc. Natl. Acad. Sci. USA {\bf 102}, 7794 (2005).

\bibitem{guimera} R.~Guimer\`a, and L.A.N. Amaral,
Eur. Phys. J. B {\bf 38}, 381-385 (2004).

\bibitem{weightbook} J.~Clark and D.A.~Holton, {\it A first look at
graph theory}, World Scientific, Second reprint 1998.

\bibitem{Barrat:2005} A.~Barrat, M.~Barth\'elemy, and A.~Vespignani, 
J. Stat. Mech. (2005) P05003. 

\bibitem{yookwt}
S.H. Yook, H. Jeong, A.-L. Barabasi and Y. Tu
Physical Review Letters {\bf 86}, 5835 (2001).

\bibitem{freeman}
L. Freeman, Sociometry,  {\bf 40}, 35 (1977).

\bibitem{num_rec} Numerical recipes in Fortran, W.H. Press,
B.P. Flannery, S.A. Teukolsky, W.T. Vetterling, Cambridge University
Press, 2nd Edition 1992.

\bibitem{holme02}
P. Holme, B. J. Kim, C. N. Yoon, S. K. Han
Phys. Rev. E {\bf 65}, 056109 (2002).

\bibitem{brandes}
U. Brandes, Journal of Math. Sociology, {\bf 25}, 35 (2001).

\bibitem{dijkstra}  
E.W.~Dijkstra, Numer. Math. {\bf 1}, 269 (1959).

\bibitem{crucitti}
V. Latora and M. Marchiori, Phys. Rev. Lett. {\bf 87}, 198701 (2001).

\end{thebibliography}
\end{document}